# Chapter 1

# CHAOS AND TIDAL CAPTURE

ROSEMARY A. MARDLING[1]
*Institute of Astronomy*
*Madingley Road, Cambridge CB3 0HA, United Kingdom*

## Abstract

We review the tidal capture mechanism for binary formation, an important process in globular cluster cores and perhaps open cluster cores. Tidal capture binaries may be the precursors for some of the low-mass X-ray binaries observed in abundance in globular clusters. They may also play an important role in globular cluster dynamics. We summarize the chaos model for tidal interaction (Mardling 1995a), and discuss how this affects our understanding of the circularization process which follows capture.

## 1.1 Introduction

Various theories for binary star formation have been suggested (some more successful than others), including fission (Durisen & Tohline 1985), fragmentation (Boss 1985), conucleation (Shu et al. 1987) and dynamical formation and tidal capture (Clarke, this volume), all of which are supposed to occur around the time of star formation. Dynamical formation (see next section) depends only on the masses of the objects and requires high protostellar densities to be effective, as are found in protostellar clusters. This is also true of stellar tidal capture, a process which becomes viable in dense stellar environments such as globular cluster cores (and perhaps open cluster cores). We will describe tidal capture in detail in the next section, but suffice to say here that it involves transferring the excess energy of unbound orbital motion to the tides of two stars which pass each other at a distance of a few stellar radii, so that a bound system (a binary) can result.

[1]Permanent address: Mathematics Department, Monash University, Melbourne, Australia, `r.mardling@sci.monash.edu.au`.

The concept of tidal capture occurred to Fabian, Pringle & Rees in 1975 after several X-ray sources were discovered which appeared to be associated with globular clusters. It was inspired by the fact that the X-ray luminosity-to-mass ratio was more than 100 times greater in globular clusters than for the galaxy as a whole (Katz 1975), so that the mechanism responsible for the X-ray sources must be peculiar to globular clusters (but see Mardling 1996). They showed that it was possible for capture to occur in a globular cluster (as opposed to only collision or flyby) and proposed that the X-ray sources were accreting neutron stars. Press & Teukolsky (1977) followed up this work by calculating the energy transfer to the tides of a polytrope of index 3 during periastron passage of a parabolic orbit (see also Lee & Ostriker 1986 and Giersz 1986). At around the same time, it was suggested by Gunn & Griffin (1979) that globular clusters were devoid of binaries. Since it was becoming clear that binaries could provide a source of energy for core support against collapse,[2] tidal capture binaries took on this role as well as the role of progenitors for the X-ray sources.

Later McMillan, McDermott & Taam (1987) suggested that the orbital circularization process following capture destroys the binary (see also Ray, Kembhavi & Antia 1987). They argued that energy transfer at periastron is one way (from orbit to tides) so that the binary must circularize very quickly (in as little as 10 yr). They estimated a dissipation timescale of $10^4 - 10^6$ yr so that by the time of circularization, a system has dissipated only a small fraction of the total tidal energy. The tidal energy will be equal to the binding energy of the binary since the binary orbit at capture is approximately parabolic. Assuming that the orbital angular momentum remains constant (*ie.* assuming there is no angular momentum loss via mass loss, gravitational radiation or magnetic breaking), the final orbital separation will be twice the periastron separation at capture. For an equal-mass system with a capture periastron separation of 3 stellar radii, the tidal energy will be 20% of the internal binding energy of the polytrope. The stars will respond by expanding; since they are so close initially, a common envelope phase will be entered into and the binary may be destroyed.

But this didn't leave globular clusters devoid of binaries; finally some giants were observed in the outer regions of a few clusters, from which a binary fraction of about 10% was inferred (Pryor et al. 1989).

In this way tidal capture fell from favour as an important process in globular clusters. But it still remained to explain the origin of the X-ray sources, long agreed to be low-mass X-ray binaries (LMXBs - neutron stars accreting from low mass main sequence or giant companions). Hut, Murphy & Verbunt (1991) suggested that such binaries were the result of clean exchanges of neutron stars into primordial binaries. However, this process tends to leave binaries with periods considerably longer than those observed (Mardling 1995b). Since the mass ratio of a neutron star to a globular cluster main sequence star (the turnoff mass is $\sim 0.8 M_\odot$) is such that a common envelope phase will not be entered into, the only obvious way to shrink the orbit within a Hubble time is through encounters with other stars. It remains to be seen whether this process is efficient enough to produce the LMXBs observed (Davies 1995).

[2]Indeed it was shown using Fokker-Planck simulations that the tidal capture binaries produced by a cluster are capable of halting core collapse and could even fuel reexpansion (Statler, Ostriker & Cohn 1986).

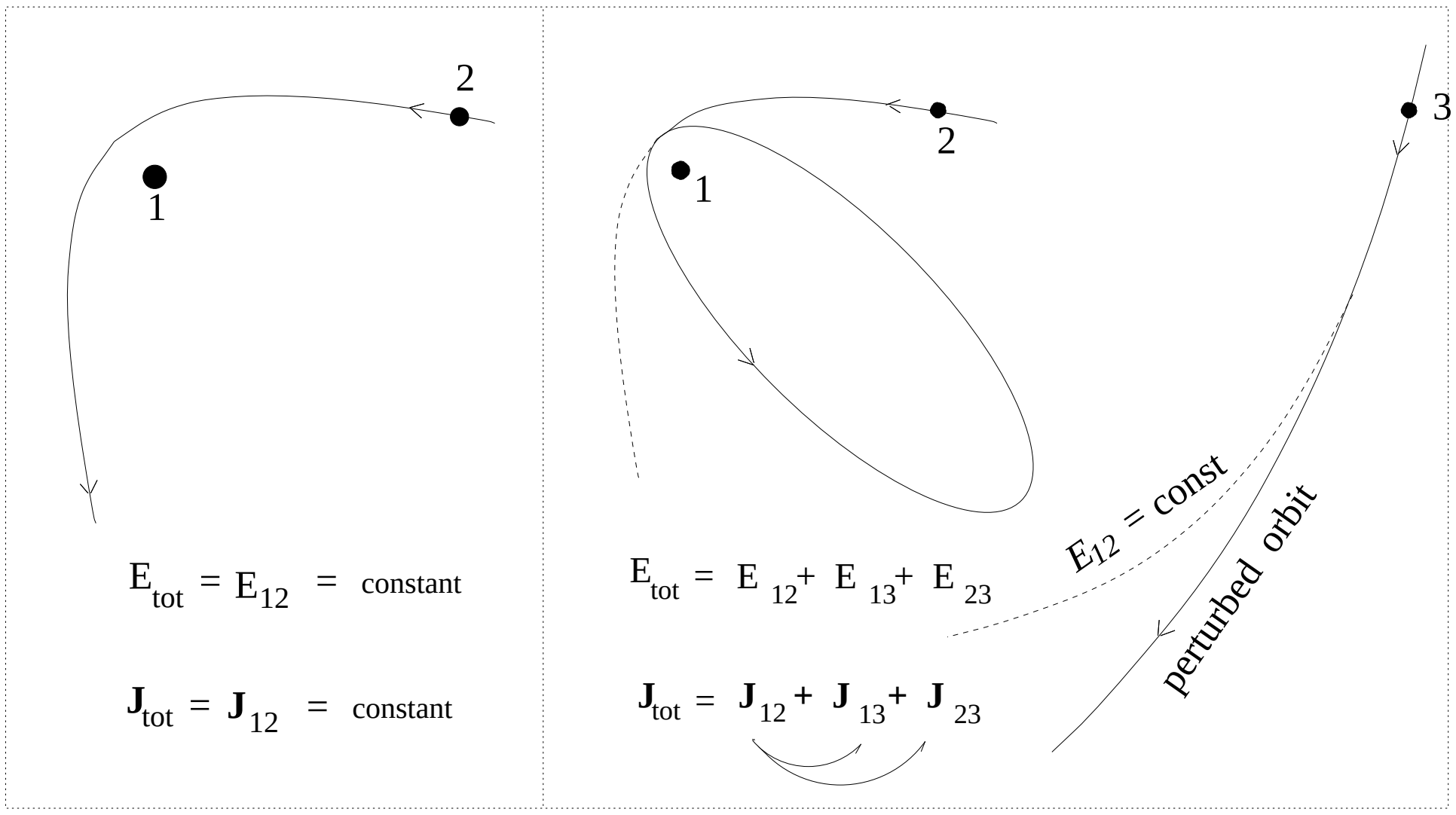

**Figure 1.1**: Dynamical formation of a binary. Here $E_{\rm tot}$ and $J_{\rm tot}$ are the total energy and angular momentum respectively, $E_{ij}$ is the orbital binding energy of the pair $(i, j)$, and $J_{ij}$ is the angular momentum about the centre of mass of the pair.

And it still remained to understand what actually happens following capture. We will address this question in the following section in which we describe in more detail the tidal capture process. Section 2 summarizes the chaos model for tidal interaction (Mardling 1995a), while section 3 applies this model to capture orbits. Section 4 discusses dissipation and long-term evolution, while the final section presents a discussion.

## 1.2 Tidal Capture

Consider an isolated system of two point masses moving in a hyperbolic orbit relative to each other. It is not possible for such a system to produce a binary; the relative orbit is fixed along with the the total energy and angular momentum. Now introduce a third body into the system. Fig. 1.1 illustrates how it is possible to produce a binary from the original pair by transferring some of its orbital energy and angular momentum to the third body. This process of forming binaries dynamically occurs in dense clusters; in fact it is *inevitable* in a cluster consisting only of point masses and which contains no binaries initially (Aarseth 1974).

Now return to two bodies but allow one of them to be finite in size and non-rigid. Let their relative speed at infinite separation be $v_\infty$ so that the total energy of the system is $\frac{1}{2}\mu v_\infty^2$, where $\mu = M_1M_2/(M_1+M_2)$ is the reduced mass with $M_1$ and $M_2$ the masses of the particles. Now it is possible for the non-rigid body to absorb some of the energy of the system; if it can manage to absorb at least $\frac{1}{2}\mu v_\infty^2$, the orbital binding energy of the pair will become negative and a binary will have been formed. This is the process of tidal capture (see fig. 1.2). Now let us examine the energy transfer process. It is easier to imagine the extended object as being composed of $N$ particles rather than a continuous fluid. Each particle experiences a force from the companion object, and this force will vary from particle to particle depending on its distance from the companion. At the same time, each particle experiences the gravitational force due to all the other particles, as well as a pressure force. As the companion draws near, the particles are pulled towards it and the fluid body assumes a distorted shape. But the restoring forces of pressure and self-gravity respond and an oscillation is set up. This process is commonly referred to as dissipative because in all but ideal circumstances, the oscillations will damp via viscous processes within the star. We refer to the oscillation that is set up as the *tides* of the star.

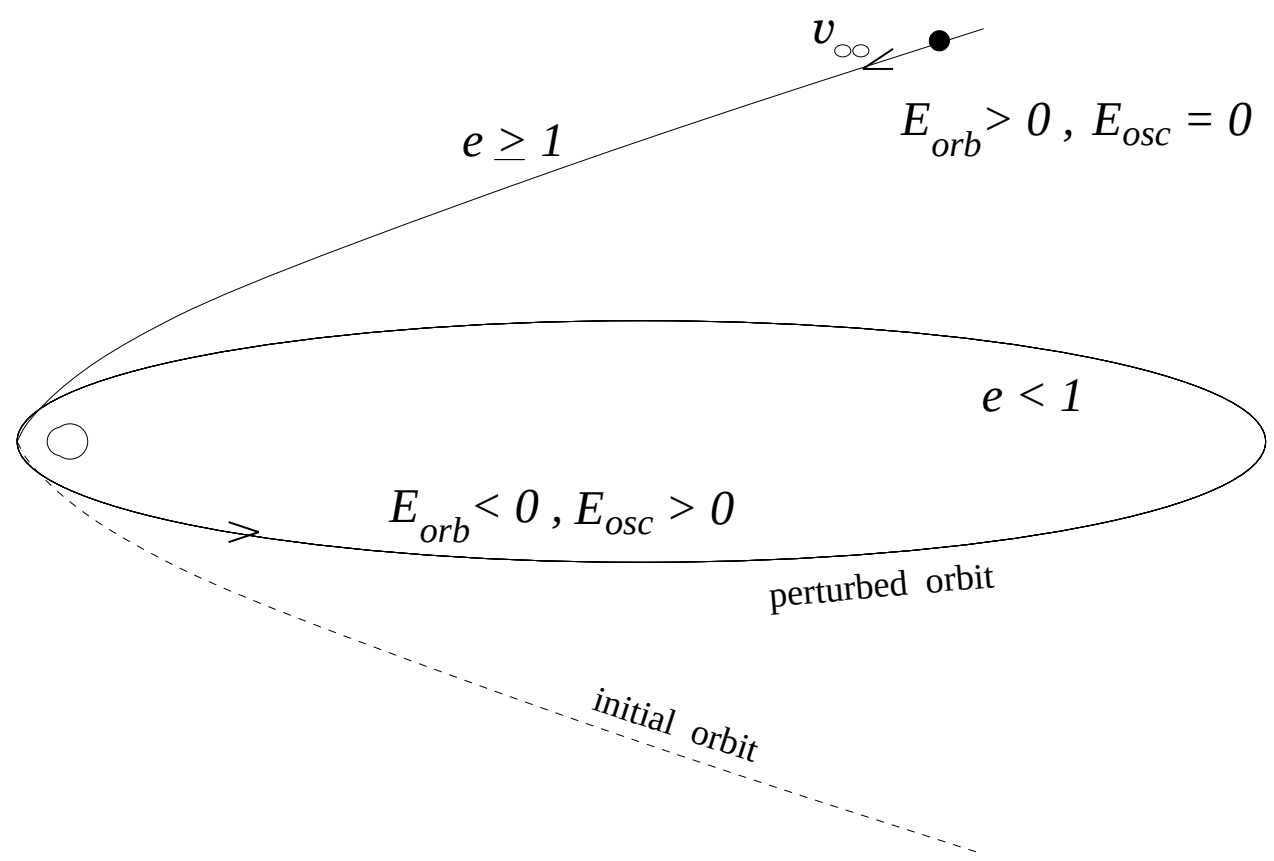

**Figure 1.2**: Tidal capture formation of a binary. $E_{\rm orb}$ and $E_{\rm osc}$ are the orbital binding energy and the tidal or oscillation energy respectively, and $e$ is the orbital eccentricity.

Just as in the case of a violin string or a drum, the oscillation can be decomposed into the object's *normal modes* of vibration. A string being 1-dimensional has one set of eigenmodes which are the circular functions. A drum has two sets; the *radial* eigenfunctions are Bessel functions, while the *azimuthal* eigenfunctions are again the trigonometric functions. The normal modes of a star will be made up of three sets of eigenfunctions. Given an equation of state for the fluid the radial eigenfunctions can be determined. These will come from a system of equations which resemble a Sturm-Liouville system. The polar angle dependence is via Legendre functions and the azimuthal dependence is trigonometric, these two together forming the spherical harmonics (Jackson 1975). The modes exited in a star by a companion are called the *non-radial* modes of vibration because they have angular dependence. Purely radial modes require a radially symmetric excitation mechanism as in the case of the Cepheid variables.

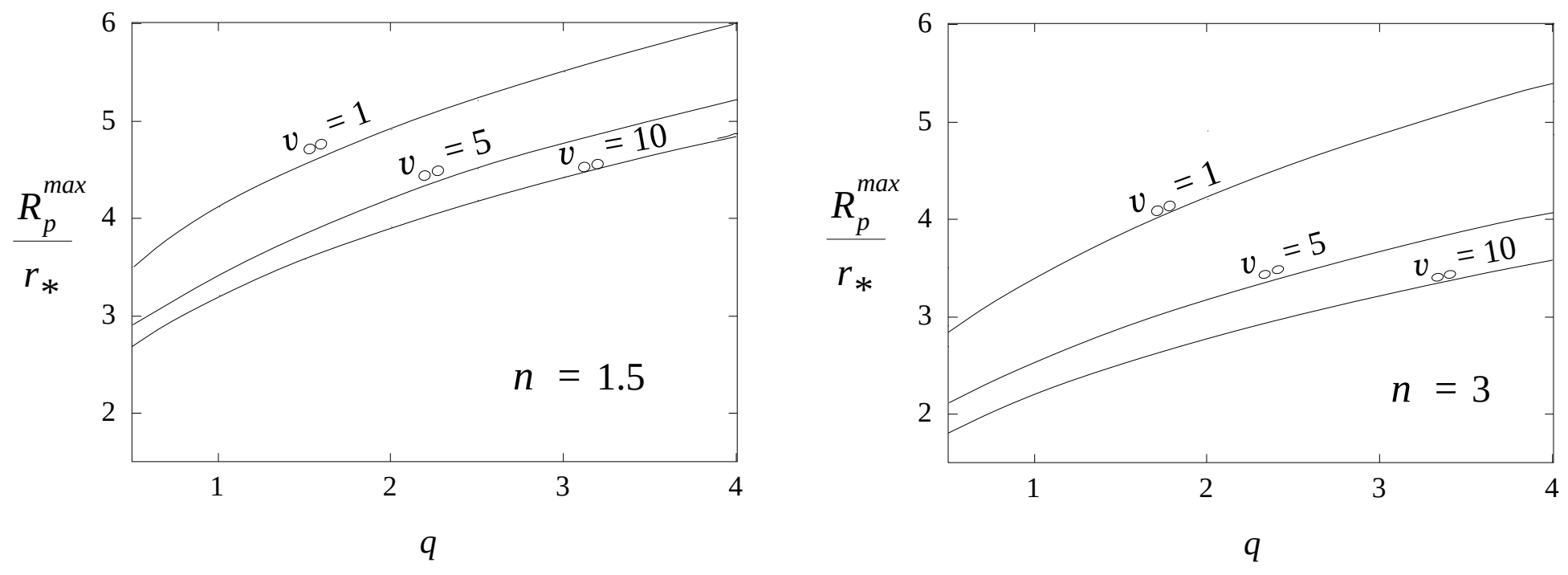

**Figure 1.3**: Maximum periastron separation for capture, $R_p^{\rm max}$ (in units of the stellar radius), versus mass ratio $q$ for point mass/polytrope models with indices $n = 1.5$ and 3. Here $q$ is the mass ratio of the point mass to the polytrope and $v_\infty$ is the relative velocity at infinity measured in km s$^{-1}$.

Whether or not a capture results in a stable binary (at least in the short term) will depend on how close the two stars come as well as the system's total energy, all of which must be absorbed by the tides. If the total energy of the system is excessive, the stars will essentially need to collide in order to transfer enough energy to create a bound system. If the stars avoid collision and manage to form a binary, they may still tidally disrupt on a subsequent periastron passage.

### 1.2.1 Capture Cross-Section

How close do stars need to come for a capture to occur? For velocity dispersions typical of open and globular clusters (1 km s$^{-1}$ and 10 km s$^{-1}$ respectively), their separation at closest approach must be at most a few stellar radii (fig. 1.3). But the process is significantly enhanced by the *gravitational focussing* of the two bodies. Fig. 1.4 illustrates how this works with two point masses. Let $R_0$ be the perpendicular distance between the two bodies at infinity (the impact parameter) and $R_p$ ($p$ for periastron) be the separation at closest approach, and let $v_p = R_p\dot{\varphi}_p$ be the velocity of the companion relative to the first star, with $\dot{\varphi}_p$ being the angular velocity at periastron. Conservation of angular momentum gives

$$R_0 v_\infty = R_p v_p \tag{1.1}$$

while the total energy of the system is given by

$$\tfrac{1}{2}\mu v_\infty^2 = \tfrac{1}{2}\mu v_p^2 - \frac{GM_1M_2}{R_p}. \tag{1.2}$$

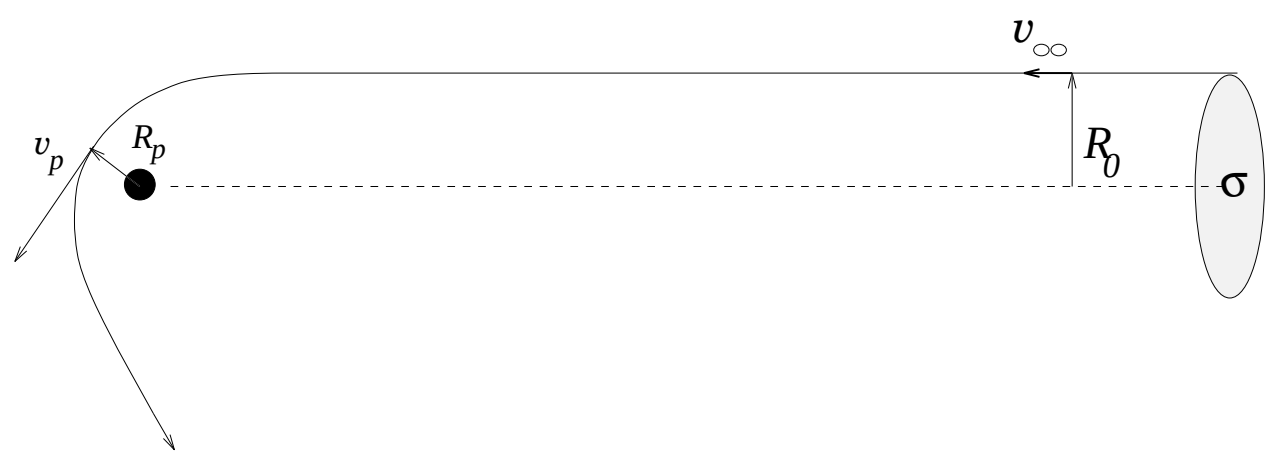

**Figure 1.4**: Gravitational focussing.

**Table 1.1**: The effectiveness of gravitational focussing

| $v_\infty$ (km s$^{-1}$) | $R_p^{\rm max}$ | $R_0$ |
|---|---|---|
| 10................ | $3.2r_*$ | 1 AU |
| 1................ | $4.1r_*$ | 12.6 AU $\sim 1.3R_{\rm Saturn}$ |

These two equations together give an expression for the *capture cross-section*, $\sigma$:

$$\sigma \equiv \pi R_0^2 = \pi R_p^2 \left(1 + (v_{\rm esc}/v_\infty)^2\right), \tag{1.3}$$

where $v_{\rm esc} = (2G(M_1 + M_2)/R_p)^{1/2}$ is the "escape velocity" of one star relative to the other at a distance $R_p$. Since this is generally much greater than $v_\infty$ for globular and open clusters,[3] the term involving it dominates and we have the result that $\sigma \sim R_p$. In contrast, the velocity dispersions in galactic nuclei are much higher so that the capture cross-section is not dominated by gravitational focussing in this case.

Table 1.1 illustrates the effectiveness of gravitational focussing for the capture of a neutron star (mass $1.4M_\odot$) by a main sequence star of mass $0.6M_\odot$ and radius $0.6R_\odot$. Here, $R_p^{\rm max}$ is the maximum distance of closest approach for which a capture can occur, a quantity which depends on the mode structure of the star (in this case for a polytrope of index 1.5). This can be calculated using the model to be described in this paper, or using the analysis devised by Press and Teukolsky (1976) to calculate the energy transfer during one periastron passage of a parabolic orbit.

### 1.2.2 Capture Rates

How often does tidal capture occur between two species of stars? This will depend on the stellar densities of both species, their relative velocities, and the capture cross-section (which itself depends on the relative velocities). Imagine we have species 1 and 2, say, a population of neutron stars of mass $M_1$, and a population of main sequence stars of mass $M_2$. For simplicity we can imagine that each neutron star-main sequence star pair is moving relative to each other with the same speed, in this case, the cluster velocity dispersion. If the stellar density of species 1 is $N_1$, then the number of captures of these stars by a star of species 2 in time $\delta t$ will be $N_1(v_\infty \delta t)\sigma$, that is, the number of stars inside a volume element of side length $v_\infty \delta t$ and base area $\sigma$. If the stellar density of species 2 is $N_2$, then the total number of captures that take place in a unit volume and unit time, that is the capture rate between species 1 and 2, is given by

$$\begin{aligned}\Gamma_{12} &= N_1 N_2 v_\infty \sigma \\ &= 2\pi G N_1 N_2 R_p (M_1 + M_2)/v_\infty \\ &= 6.8 \frac{N_1}{10^4 \text{ pc}^{-3}} \frac{N_2}{10^4 \text{ pc}^{-3}} \frac{M_1 + M_2}{M_\odot} \frac{R_p^{\min}}{R_\odot} \frac{10 \text{ km s}^{-1}}{v_\infty} \text{Gyr}^{-1}\text{pc}^{-3}. \quad (1.4)\end{aligned}$$

[3]For example, the escape velocity at the solar surface is 619 km s$^{-1}$.

Using data from Davies & Benz (1995) for the globular cluster 47 Tuc, the density of neutron stars in the core is 900 pc$^{-3}$, the density of turnoff mass main sequence stars is $3 \times 10^4$ pc$^{-3}$ and the velocity dispersion is 10 km s$^{-1}$ so that the capture rate is 12 captures Gyr$^{-1}$ pc$^{-3}$.

### 1.2.3 Evolution Following Capture

What really happens following capture? The energy transfer during the first periastron passage is well established, at least for non-rotating non-pulsating stars in parabolic orbits. But the energy transfer during subsequent passages will depend on whether or not the stars are still oscillating, which in turn will depend on the dissipation timescales of the stars compared with the orbital periods. For captures occurring with velocities at infinity typical of globular cluster (3-D) velocity dispersions (around 10 km s$^{-1}$), the orbital periods are less than the estimated *linear* dissipation timescale (see fig. 1.5), while for lower velocities (such as would be found in *open* clusters), the situation can reverse. On the other hand, for very energetic tides *nonlinear* effects can become important and the dissipation timescale may be much shorter. But if the stars *are* still oscillating during a periastron passage, the *direction* of energy flow will depend on the *phase* of oscillation; it is possible for energy to be returned to the orbit (Mardling 1991, Kochanek 1992). Thus it is also possible for a newly formed binary to *self-ionize* if the system has not yet dissipated all the energy of the initial unbound orbit. But it can be shown that the majority of systems avoid this fate (Mardling 1995b).

The energy transfer at periastron will also depend on the tidal energy already present. In fact as we will show in section 1.3, this energy transfer process is *chaotic*; the evolution of the orbit following capture depends sensitively on conditions at capture, at least until the system has dissipated an amount of energy which depends again on conditions at capture.

### 1.2.4 Dynamical Calculations

In order to study the dynamical evolution following capture, one must follow in detail the individual oscillations which occur on the dynamical timescale of the star. But the orbital periods immediately after capture are extremely long. For example, for an equal-mass capture at 3 stellar radii with a velocity at infinity of 10 km $s^{-1}$ (corresponding to an eccentricity of 1.0008 before capture and 0.999 following capture), the orbital period is $1.7 \times 10^5$ oscillations of the lowest order (most energetic) mode. This fact has previously prevented the dynamical calculation of all but a few orbits following capture.

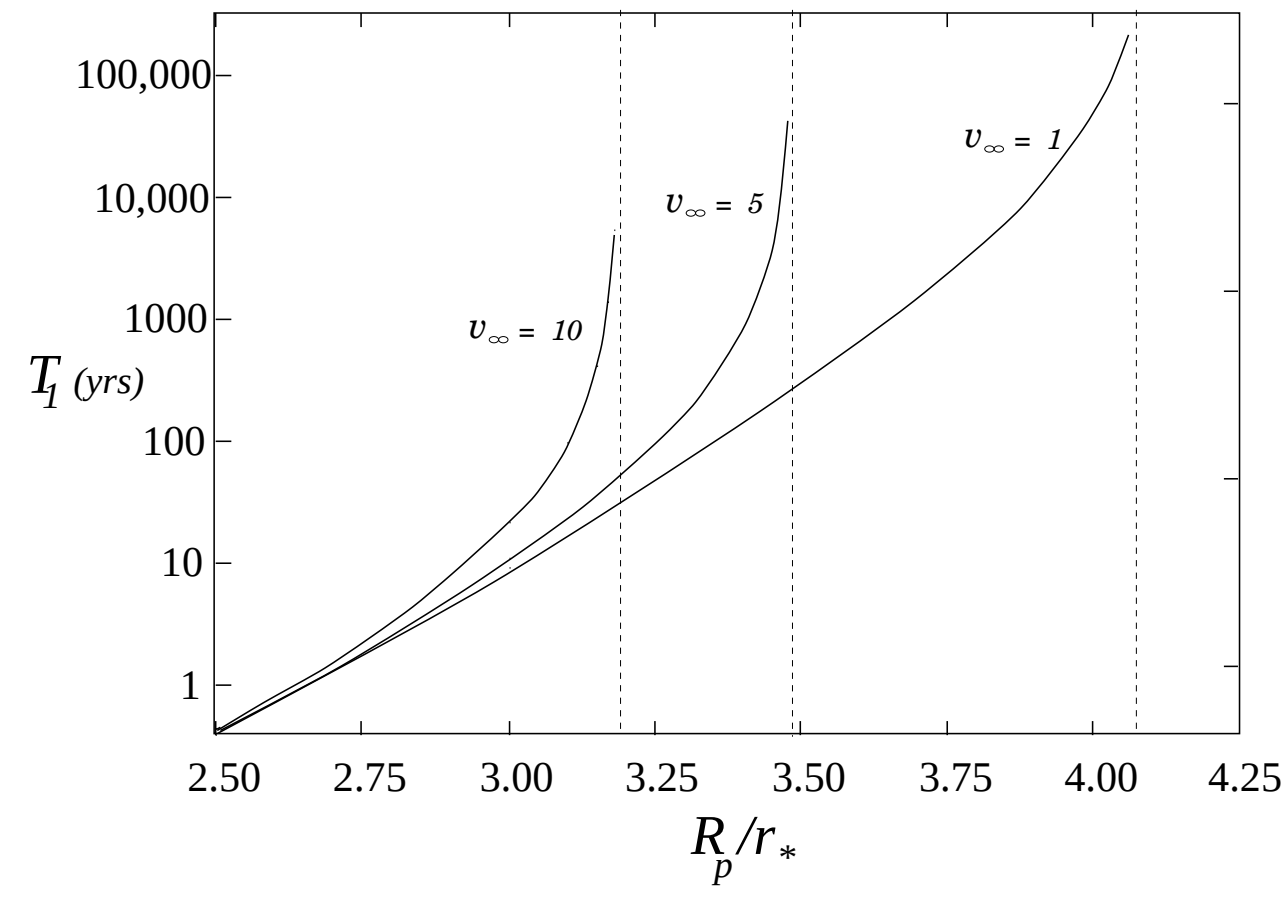

**Figure 1.5**: The first orbital period following capture as a function of periastron separation at capture for various velocities at infinity (measured in km $s^{-1}$). The asymptotes indicate the maximum periastron separation for which capture is possible.

Several numerical methods have been applied. Smoothed particle hydrodynamics (SPH) was used to study collisions and extremely close encounters (see for example Gingold & Monaghan 1980, Rasio & Shapiro 1991 and Benz & Hills 1992).

A method devised by Carter & Luminet (1985) which employs a polytropic version of Chandrasekhar's homogeneous ellipsoid was used by Kochanek (1992) to model the first few orbits following capture.

A third method is the subject of the rest of this review. It is a modification of a *normal mode analysis* devised by Gingold & Monaghan (1980). The success of this method lies with the fact that the equations of motion are derived from a Lagrangian so that the total energy (and angular momentum) are conserved and energy is free to flow between the orbit and the tides. Details may be found in Mardling (1995a,b); here we present a summary.

## 1.3 The Chaos Model for Tidal Interaction

We now examine more formally the energy transfer process of the tidal interaction between two stars (here modelled as a point mass and a non-rotating polytrope). It is possible to construct a Lagrangian consisting of terms corresponding to a Kepler orbit, terms corresponding to a self-gravitating fluid and a term corresponding to the interaction between the two. Thus energy is free to flow either way between the orbit and the tides, and this interaction can be followed in detail. This calculation can be found in Gingold & Monaghan (1980) and Mardling (1995a); here we present an outline of how the method works.[4]

Given a polytropic equation of state, one then assumes small perturbations to the density and velocity field of the polytrope. The perturbation variables satisfy normal mode equations as discussed above. The *Lagrangian* is then expanded to second order in the perturbation variables and orthogonality conditions associated with the normal modes are used to perform any integrals over the polytrope. In this way, a Lagrangian is derived which depends on the orbital variables as well as the time-dependent mode amplitudes. One can then derive equations of motion for these variables and a conserved energy (and angular momentum). Higher order perturbation expansions can also be done in this way[5] (Mardling 1991), and since the equations of motion are derived from a Lagrangian, one is always assured of the existence of a conserved energy.

The equations of motion are in the following form:

$$\ddot{b}_{\mathbf{k}} + \omega_{kl}^2 b_{\mathbf{k}} = q C_{\mathbf{k}}^{(1)} \frac{\mathrm{e}^{-im\varphi}}{R^{l+1}}, \tag{1.5}$$

$$\mu \ddot{\mathbf{R}} = -\frac{GM_1M_2}{R^2}\hat{\mathbf{R}} + q\sum_{\mathbf{k}} C_{\mathbf{k}}^{(2)} b_{\mathbf{k}} \frac{\partial}{\partial \mathbf{R}}\left(\frac{\mathrm{e}^{im\varphi}}{R^{l+1}}\right), \tag{1.6}$$

where $b_{\mathbf{k}}$ is the (complex) amplitude of the mode with mode numbers $\mathbf{k} \equiv klm$, $\sum_{\mathbf{k}} \equiv \sum_{k=1}^{\infty}\sum_{l=2}^{\infty}\sum_{m=-l}^{l}$, $\omega_{kl}$ is the frequency of mode $\mathbf{k}$, $\varphi$ is the true anomoly measured from the initial line of apses, $R = |\mathbf{R}|$ is the binary separation, $q$ is the mass ratio of the point mass to the polytrope, and $C_{\mathbf{k}}^{(1)}$ and $C_{\mathbf{k}}^{(2)}$ are constants which depend on the internal structure of the polytrope. Eq. (1.5) represents a forced harmonic oscillator, the forcing being provided by the orbit, and eq. (1.6) is a Kepler orbit with a mode-orbit coupling term as a perturbation. The strength of the interaction between the orbit and the tides increases as the central condensation of the polytrope decreases, or as the mass ratio is increased. The solution is constrained by an energy integral as well as an angular momentum integral:

$$E = \tfrac{1}{2}\mu\dot{\mathbf{R}}^2 - \frac{GM_1M_2}{R} - q\sum_{\mathbf{k}} C_{\mathbf{k}}^{(2)} b_{\mathbf{k}} \frac{\mathrm{e}^{im\varphi}}{R^{l+1}} + \tfrac{1}{2}\sum_{\mathbf{k}} C_{\mathbf{k}}^{(3)}(\dot{b}_{\mathbf{k}}\dot{b}_{\mathbf{k}}^* + \omega_{\mathbf{k}}^2 b_{\mathbf{k}} b_{\mathbf{k}}^*), \tag{1.7}$$

[4]Another way to set up this problem (Mardling 1991) is to regard the fluid as being composed of $N$ particles and to consider the gravitational force between each of the particles, including the point mass, and the pressure force at each particle. The continuous limit is then taken. This way of approaching a self-gravitating problem can have many advantages, especially if the problem has odd geometry; in fact the method of SPH is set up in this way (without the continuous limit).

[5]Although one must be careful to avoid divergent series by allowing the frequencies of vibration to vary in time.

$$J = \mu R^2 \dot{\varphi} + \sum_{\mathbf{k}} imC_{\mathbf{k}}^{(3)} b_{\mathbf{k}} \dot{b}_{\mathbf{k}}^*, \tag{1.8}$$

where $C_{\mathbf{k}}^{(3)} = C_{\mathbf{k}}^{(2)}/C_{\mathbf{k}}^{(1)}$ and $i = \sqrt{-1}$. The total energy $E$ consists of the orbital binding energy (the first two terms), the oscillation energy of the tides (the last two terms), and the interaction energy between the two. Note that angular momentum can be transferred to the tides *in the absence of friction*. It is commonly believed that angular momentum transfer is only possible when dissipation causes a *tidal lag*, that is, the tidal bulge of the *equilibrium tide* does not point in the direction of the companion. This in turn places a torque on the star, *ie.*, angular momentum is transferred from the orbit to the tides. But the asymmetric nature of the *dynamical tide* allows angular momentum to be transferred in the absence of dissipation.

The self-consistent system of equations (1.5) and (1.6) allows energy and angular momentum to be freely exchanged between the orbit and the tides, the consequences of which can be severe.

### 1.3.1 Chaotic and Periodic Solutions

Two types of solutions exist to the system of eqs (1.5) and (1.6). A convenient way of displaying these solutions is to plot the orbital eccentricity against the number of periastron passages. This is done in fig. 1.6 for three sets of initial conditions, with each orbit starting with an eccentricity of 0.8 (the polytropic index is 1.5). The widest orbit (a) has an initial periastron separation[6] of $3.2r_*$, where $r_*$ is the radius of the polytrope. The change in eccentricity is small from orbit to orbit, and a beating is evident in the enlargement shown at the bottom of the figure. Orbits (b) and (c) differ in their initial periastron separation, $R_p$, by less than one part in $10^5$ ($R_p = 2.9r_*$ and $R_p = 2.90001r_*$). They exhibit extreme sensitivity to initial conditions indicating chaos.

Why is the behaviour of chaotic and periodic systems so different? There appears to be no lower bound for the eccentricity of chaotic orbits (in fact there *does* exist such a lower bound; see section 1.4), while for periodic orbits, the maximum change in eccentricity is of the order of the change during one periastron passage.[7] Large changes in eccentricity imply very energetic tides which in turn imply a need to consider nonlinear mode couplings, something this model neglects (but see section 1.4). In contrast, the tidal interaction in a periodic orbit will generally raise small tides. The difference between the two behaviours can be understood mathematically from the point of view of stability. The author has devised a 3-D mapping which mimics the behaviour seen here. The system appears to follow the "intermittency route to chaos" (Pomeau & Manneville 1980). If one calculates the eigenvalues of the mapping, stability requires their absolute values to be unity (the system is non-dissipative). When a system goes unstable, *ie.* becomes chaotic, the eigenvalues wander arbitrarily far from unity. The details of this work will be published elsewhere.

[6]The orbit is actually started at apastron and the periastron separation is supplied which would be reached if the system consisted of two point masses.

[7]Except for *resonant* orbits - those for which the orbital period is nearly a multiple of the period of oscillation of the most energetic mode.

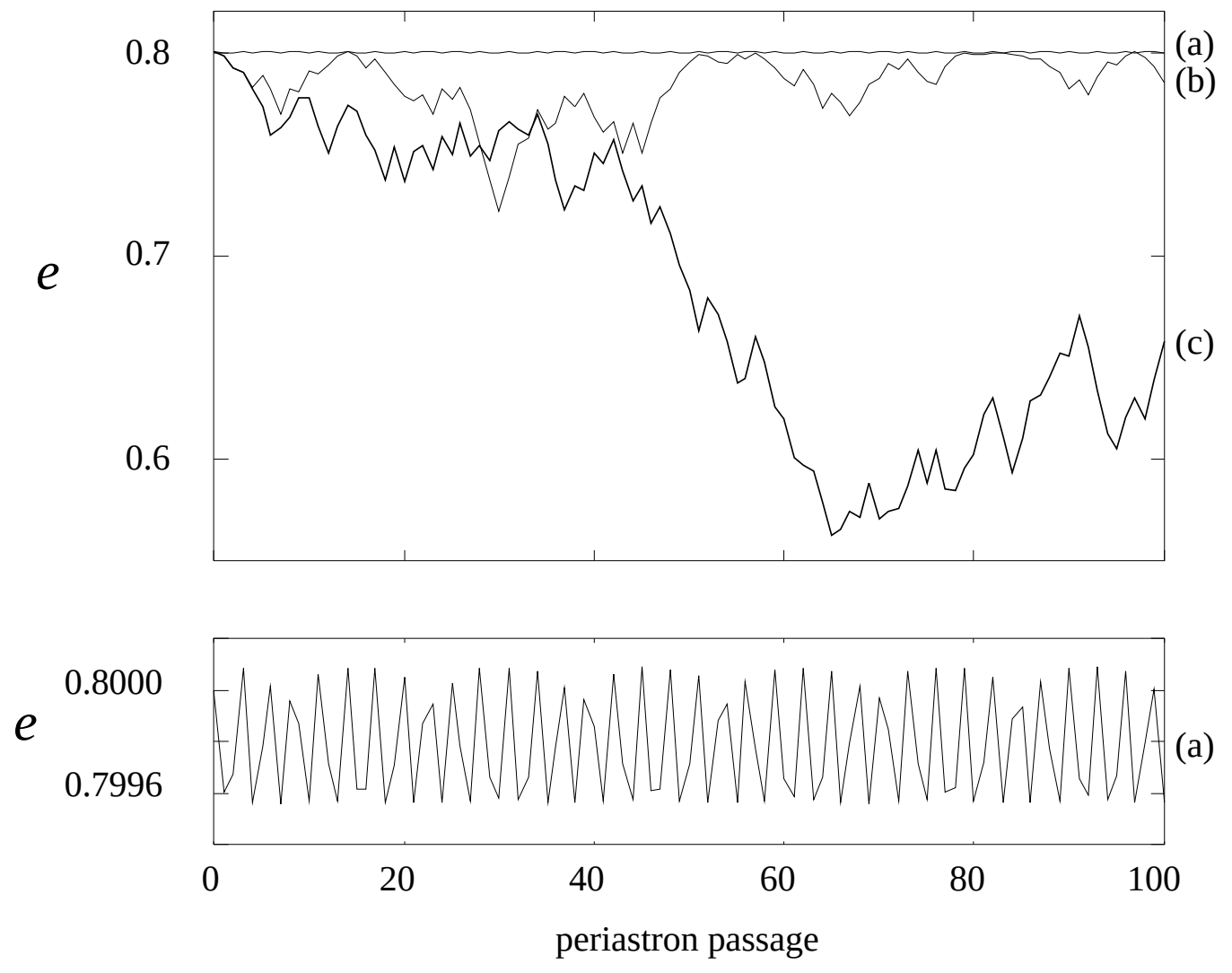

**Figure 1.6**: Chaos vs. periodicity. Curve (a) (shown magnified at the bottom of the figure) is typical of a periodic orbit for which the eccentricity varies little so that the tidal energy is small. In contrast, curves (b) and (c) display the extreme sensitivity to initial conditions typical of chaotic systems, with the eccentricity varying over a wide range and the tidal energy becoming very large.

### 1.3.2 The Chaos Boundary

The behaviour demonstrated in fig. 1.6 is typical of chaotic and periodic systems. We can discover the range of initial periastron separations and eccentricities for which chaotic motion exists by exploiting the extreme sensitivity to initial conditions. Fig. 1.7 plots the boundary between chaotic and periodic motion for equal-mass systems containing a polytrope of index 1.5. This boundary was derived by systematically running though $(R_p, e)$ parameter space and comparing 50 periastron passages of two orbits whose initial eccentricity differs by $10^{-5}$ and whose initial tidal energy is zero. Naturally there exists the possibility that a system will appear periodic for more than 50 orbits and suddenly become chaotic. This is characteristic of systems which follow the intermittency route to chaos. Nonetheless, this kind of behaviour is restricted to regions close to the chaos boundary so that although the boundary as plotted is not well defined, it serves the purpose of delineating the two types of behaviour.

The position of the boundary depends on the mass ratio as well as the stellar structure. If we define the mass ratio to be the mass of the compact object to the mass of the polytrope, then increasing this ratio causes the chaos boundary to move up in $(R_p, e)$ space. This reflects the fact that for a fixed separation, the tidal force increases as the mass ratio is increased. This can also be seen in eqs. (1.5) and (1.6); the strength of the interaction is proportional to the mass ratio. Similarly, the chaos boundary moves down if the central condensation of the polytrope increases (equivalently, if the polytropic index increases). For a fixed stellar radius and stellar separation, the energy the tides are capable of storing decreases with increasing central condensation. This reflects the mode structure which in turn reflects the mass distribution.

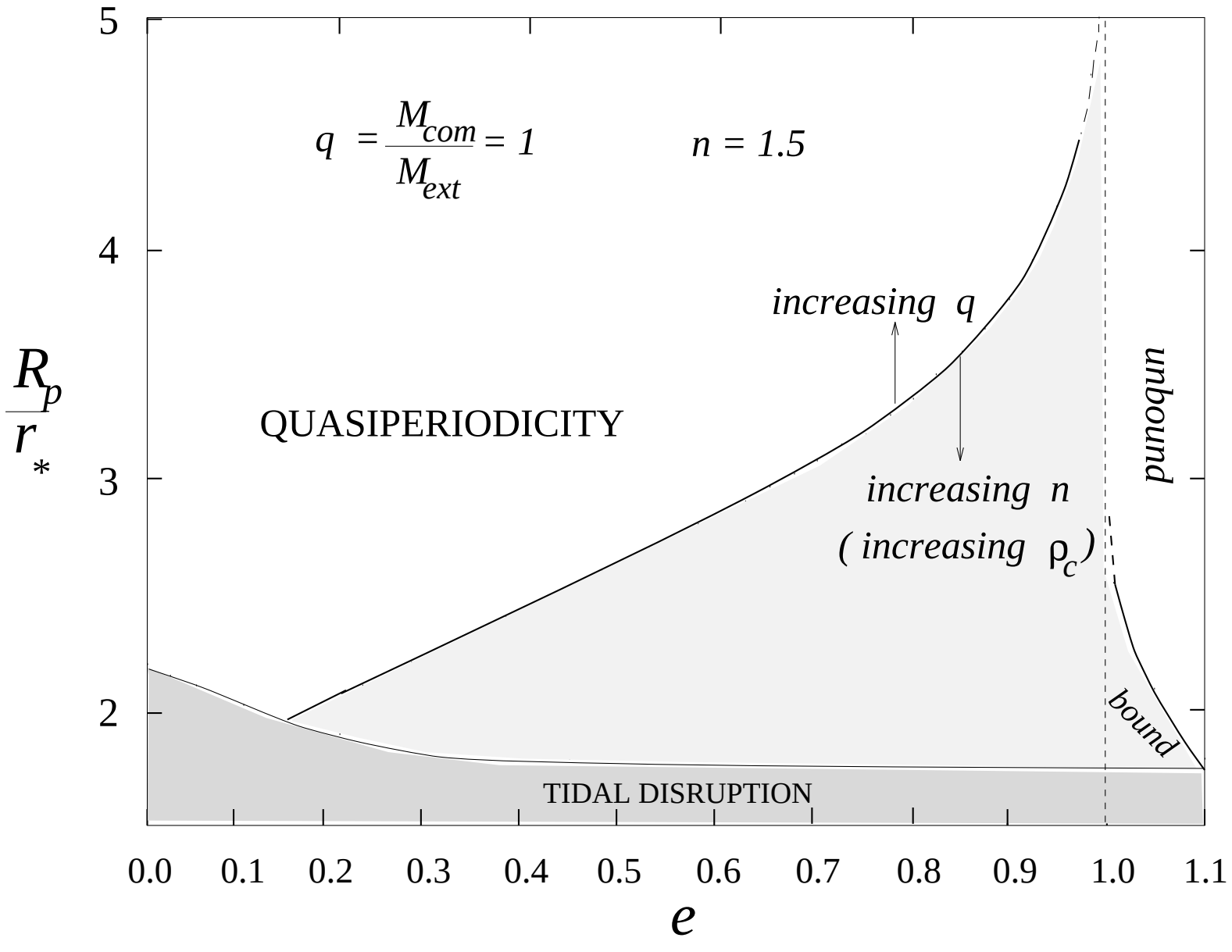

**Figure 1.7**: The chaos boundary. The polytropic index is 1.5 and the mass ratio of compact to extended star is unity. The boundary moves up if the mass ratio is increased, and down if the polytropic index (central condensation) is increased. The boundary for tidal disruption is underestimated by this model.

The chaos boundary is not very sensitive to the following:

1. The number of modes in the calculation. Including more than the $l = 2$ $f$-mode ($k = 1$) has very little effect on the position of the chaos boundary.

2. A finite radius companion.

3. Including mode-mode interactions (*ie.* including higher order terms in the perturbation expansion).

Each case can be understood in terms of stability as discussed above; the eigenvalues of the system are not much affected by these factors.

An interesting feature in fig. 1.7 is the asymptote at $e = 1$: all (nondissipative) capture binaries are chaotic. The asymptotic nature of the chaos boundary can also be understood in terms of the mapping representation of this problem (Mardling, in preparation). As well there exists a second branch to the chaos boundary which corresponds to hyperbolic capture orbits.

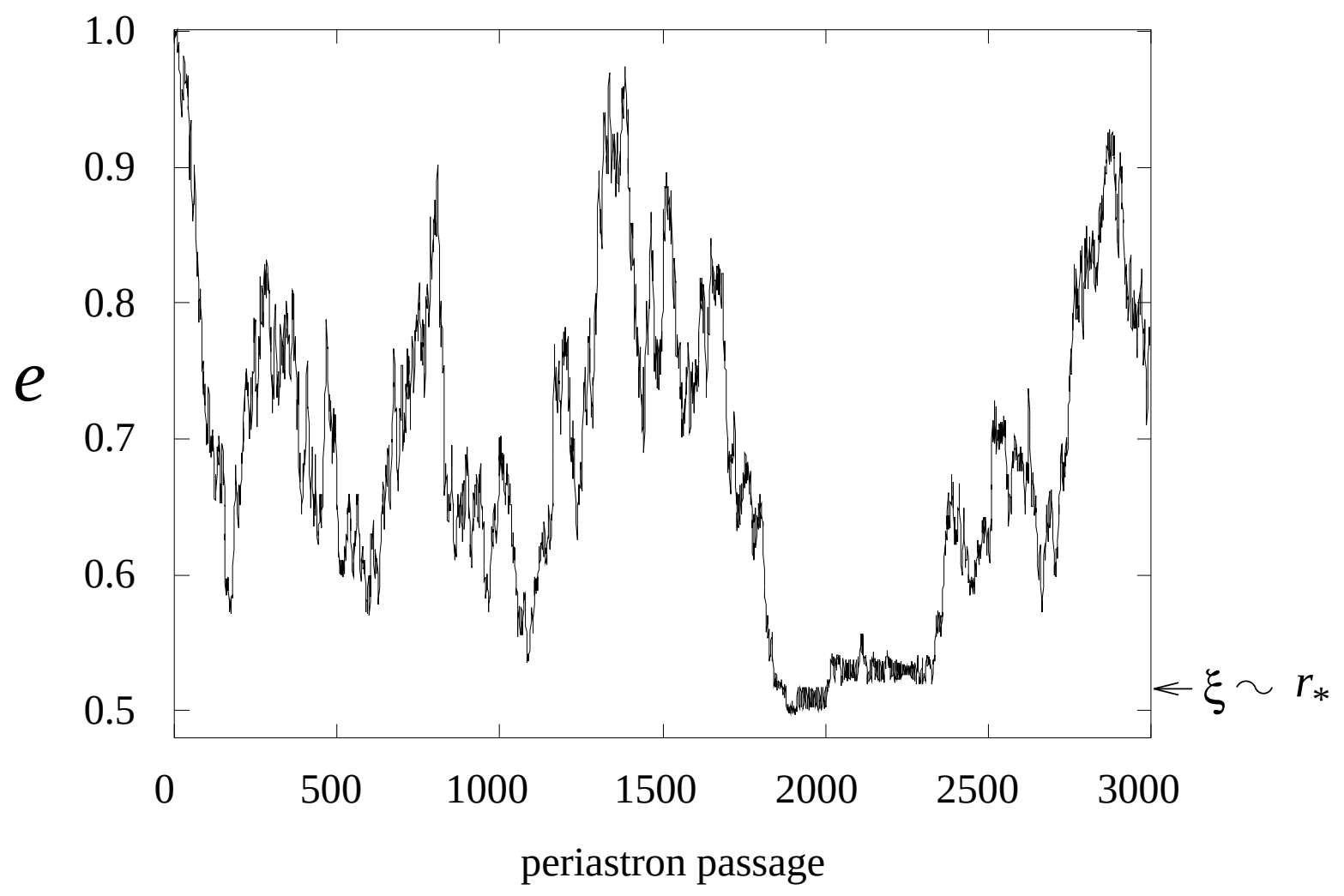

**Figure 1.8**: A capture orbit with an initial eccentricity of unity and a periastron separation at capture of $3r_*$ (no dissipation). Note the periodic feature around the 2000th orbit. At this point, the height $\xi$ of the tides is of the order of the stellar radius.

## 1.4 Capture Orbits

We now apply the chaos model for tidal interaction to the problem of tidal capture. We start by considering a dissipationless model, a model which is highly unrealistic because dissipation plays a major role in the dynamical evolution following capture. Nonetheless, this enables us to examine certain properties of such systems.

But first we must overcome the problem of computing very long period orbits while at the same time following the motion of the tides. We do this by recognizing that the tides and the orbit essentially only interact for a fraction of a very long period orbit - that is for that section of the orbit near periastron. We thus calculate this section by numerically solving eqs (1.5) and (1.6) sufficiently far past periastron so that the total energy and angular momentum (eqs. (1.7) and (1.8)) are conserved to within some tolerance[8] after several thousand orbits. The rest of the orbit is solved for analytically (see Mardling 1995b).

Figure 1.8 shows 3000 orbits of an equal-mass system in an initially parabolic orbit with an initial periastron separation of $3r_*$ (the calculation includes up to the $l = 4$ mode for $k = 1$ only). The eccentricity varies over a wide range but there exists a lower limit, in this case $e = 0.5$. The quasiperiodic behaviour seen around the 2000th orbit indicates that the system has met a chaos boundary.[9] *In the absence of dissipation, the eccentricity cannot drop below this value.*

[8]To within 1 part in $10^4$ in the example which follows. This corresponds to halting the numerical calculation at a stellar separation of about 40 stellar radii.

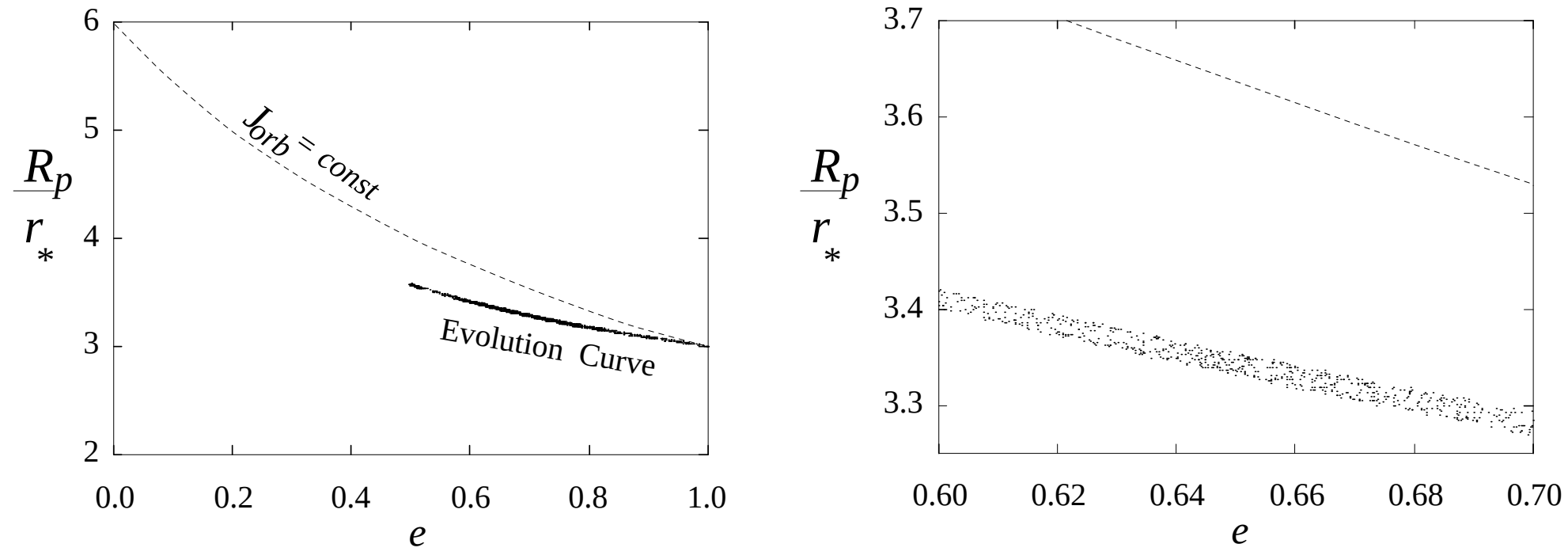

**Figure 1.9**: The evolution curve. Comparison with the curve along which the orbital angular momentum is constant indicates that a significant amount of angular momentum is transferred to the polytrope. Note the endpoint of the evolution curve which represents a minimum orbital eccentricity the system can achieve in the absence of friction. The finite width of this curve is mainly due to the influence of the $l = 3$ mode and relects its independence from the $l = 2$ mode.

Since the tides are extremely energetic (this calculation implies a tidal height of the order of the stellar radius when $e = 0.5$), a linear analysis is clearly inadequate. We argue though, that neglecting nonlinear mode-mode interactions is equivalent to neglecting a strong source of dissipation, at least for systems sufficiently far from tidal disruption. Since the orbit can only drive the low order modes (the tidal energy in a mode is proportional to $R^{-(2l+2)}$) and since the system is chaotic, it is impossible to distinguish between the present model and a model which includes mode-mode interactions (Mardling, in preparation).

It is also instructive to plot $R_p$ against $e$ and compare this *evolution curve* with the curve along which the orbital angular momentum is constant, the latter being given by $R_p(1+e) = R_p^0(1+e_0)$, with $(R_p^0, e_0)$ the initial orbital variables at capture. Fig. (1.9) shows this comparison along with an enlargement of a portion of the figure, showing that the evolution "curve" is actually of finite width and consists of points which wander up and down in a chaotic manner. The difference in the two curves reflects the angular momentum transferred to the tides, despite the system being frictionless. In fact, one can show that the angular momentum in a particular mode is proportional to the energy in that mode.

The left-hand end of the evolution curve defines the minimum possible eccen-tricity the system is capable of attaining (in the absence of dissipation),[10] which in turn limits the maximum possible tidal energy. In this example, the maximum possible tidal energy is 7% of the stellar internal binding energy,[11] quite a large fraction (although nonlinear dissipation is likely to severely limit this maximum - see the next section), while for wider systems, this maximum can be arbitrarily small. We will see in the next section that this has important implications for the survival of binaries after capture.

[9] A non-zero tidal energy chaos boundary; see Mardling (1995a) for details.

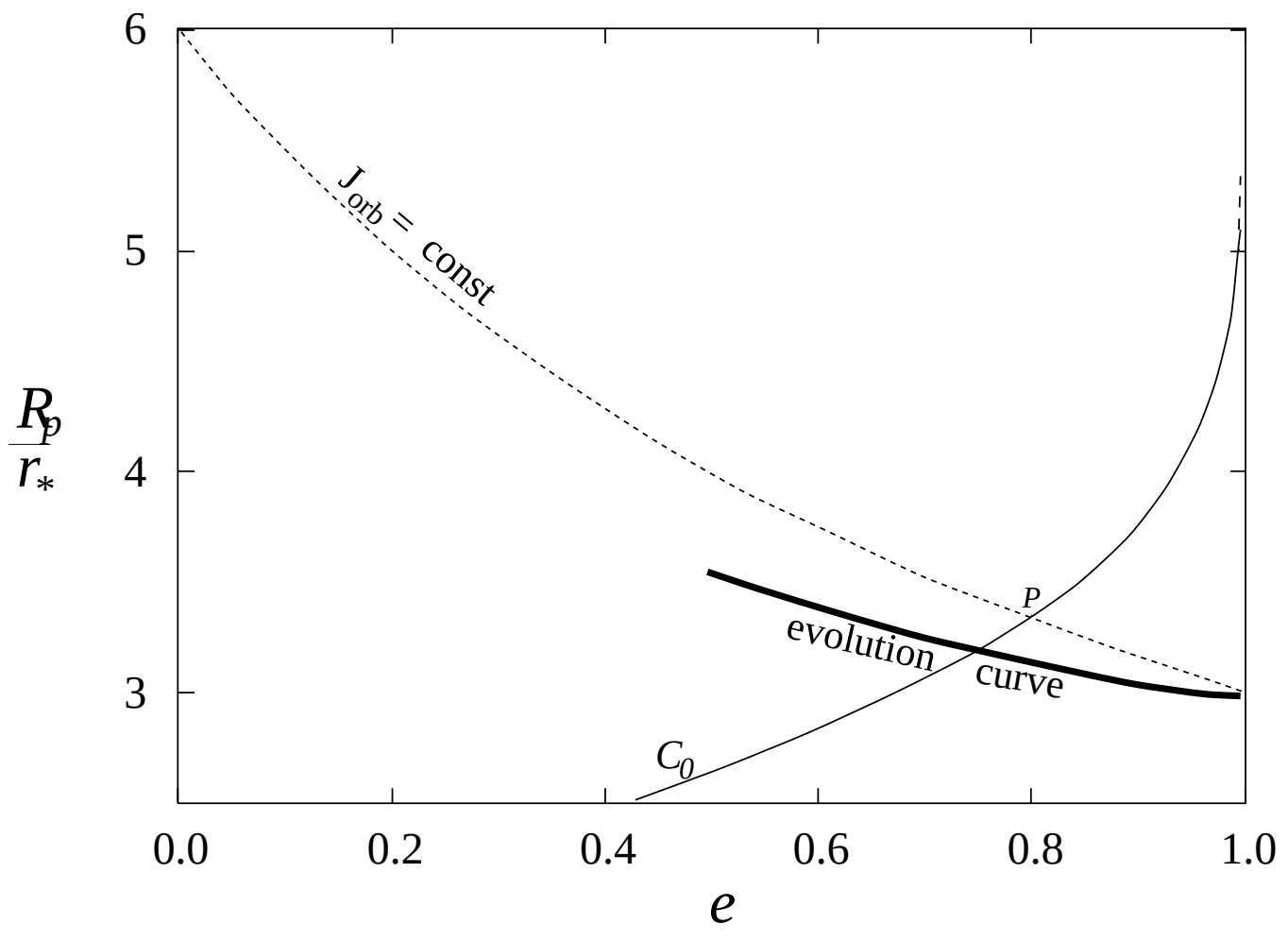

**Figure 1.10**: The effect of dissipation: the evolution curve shrinks towards the point $P$ after which the system becomes permanently periodic and evolves along the curve of constant angular momentum.

## 1.5 Dissipation and the Long-Term Evolution

We have neglected dissipation so far in order to demonstrate the chaotic nature of capture systems. Assuming there is no mass loss or transfer, and that the radii of the stars remains constant, it is possible to examine qualitatively the effect of dissipation on the orbital dynamics without knowing the details of the dissipation mechanisms. Consider the effect on the evolution curve as the system loses energy. The *maximum* possible eccentricity will decrease as the system becomes more bound, and since this corresponds to zero tidal energy, the endpoint of the evolution curve must sit on the curve of constant orbital angular momentum, $J_{\rm orb} = \rm const$ (fig. 1.10). At the same time, the *minimum* possible eccentricity will *increase*, a process which can be partially understood as follows (but see Mardling 1995b). Imagine that a system has evolved to the state in which the right hand end of the evolution curve as described above lies at the point where the chaos boundary $C_0$ intersects $J_{\rm orb}$ = const (the point $P$ in Figure 1.10). By definition, the right hand end point of the evolution curve corresponds to zero tidal energy. Since any system which starts to the left of the chaos boundary with zero tidal energy is periodic in nature, it follows that the point $P$ marks the end of the chaotic phase and that before this, the minimum eccentricity that the system can achieve at any time must increase. After the system has reached the point $P$ and is no longer chaotic, the tides will be small and hence the dissipation rate will be greatly reduced. The system will then move slowly along $J_{\rm orb}$ = const until the binary is circularized with a separation of twice the periastron separation at capture.

[10]It is possible to calculate this minimum; see Mardling (1995b).

[11]This should be compared with a figure of 20% for complete circularization had the tidal energy not been restricted by the chaos boundary

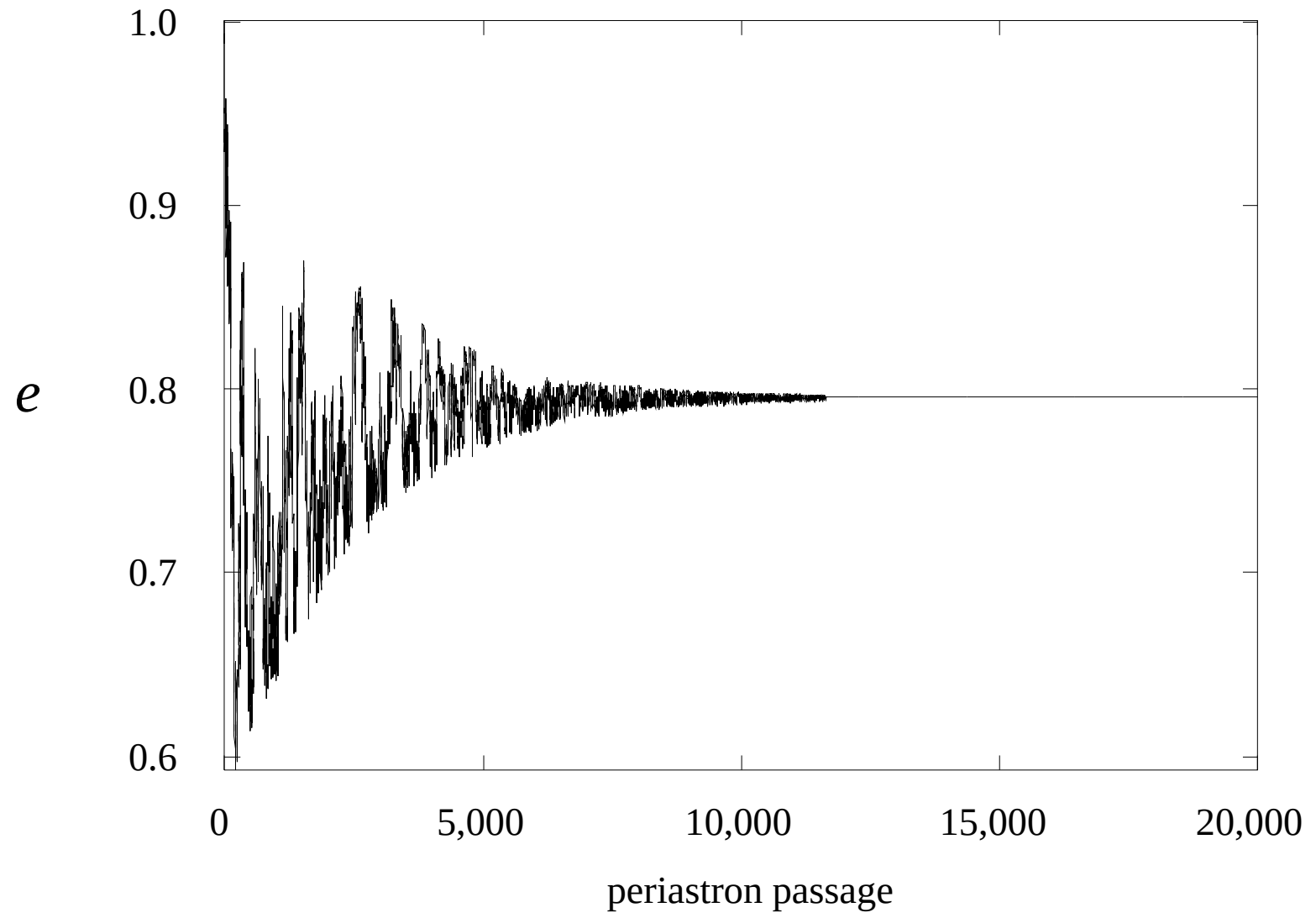

**Figure 1.11**: A capture orbit with dissipation.

This process can be modelled by introducing an artificial damping term into the equation of motion for the mode amplitudes. Eq. (1.5) is replaced by

$$\ddot{b}_{\mathbf{k}} + 2\mathcal{C}_{\mathbf{k}}\dot{b}_{\mathbf{k}} + \omega_{kl}^2 b_{\mathbf{k}} = qC_{\mathbf{k}}^{(1)}\frac{\mathrm{e}^{-im\varphi}}{R^{l+1}}, \tag{1.9}$$

where $\mathcal{C}_{\mathbf{k}} \propto 1/\tau_{\mathbf{k}}$, with $\tau_{\mathbf{k}}$ being the damping timescale for mode $\mathbf{k}$. In fact, when the tides are large, *nonlinear* damping due to mode-mode interactions will dominate *linear* damping, the latter being due to normal viscous processes. The damping timescale for mode $\mathbf{k}$ will thus be given by $1/\tau_{\mathbf{k}} = 1/\tau_{\mathbf{k}}^l + 1/\tau_{\mathbf{k}}^{nl}$, where $\tau_{\mathbf{k}}^l$ and $\tau_{\mathbf{k}}^{nl}$ are the linear and nonlinear damping timescales respectively. Figures 1.11 and 1.12 illustrate the circularization process again for an equal-mass system with initial periastron separation $R_p = 3r_*$ and initial eccentricity $e = 1$, but with a nonlinear damping timescale of about 30 yr and a linear damping timescale of 1000 yr. The nonlinear damping timescale may well be shorter than this, but whenever it is longer than the orbital period, chaotic behaviour will persist. In this case, the chaotic phase lasts for about 200 years after which time the binary takes more than 5.5 million years to finish circularizing. This represents on the order of $2 \times 10^9$ orbital periods which would take around 1000 CPU *years* to run on a fast workstation. Instead this calculation was done using a method devised by the author for use in $N$-body cluster calculations (see Aarseth 1996 for a description of the astrophysical processes included in this model). This method assumes an impulse approximation for energy transfer at periastron during the chaotic phase. The energy transfer depends on the tidal energy present as well as the oscillatory phase and uses a modified Press and Teukolsky (1977) analysis (one for which the similarity variable depends also on the orbital eccentricity). The periodic phase is calculated semi-analytically. A thorough description of the method will be published elsewhere.

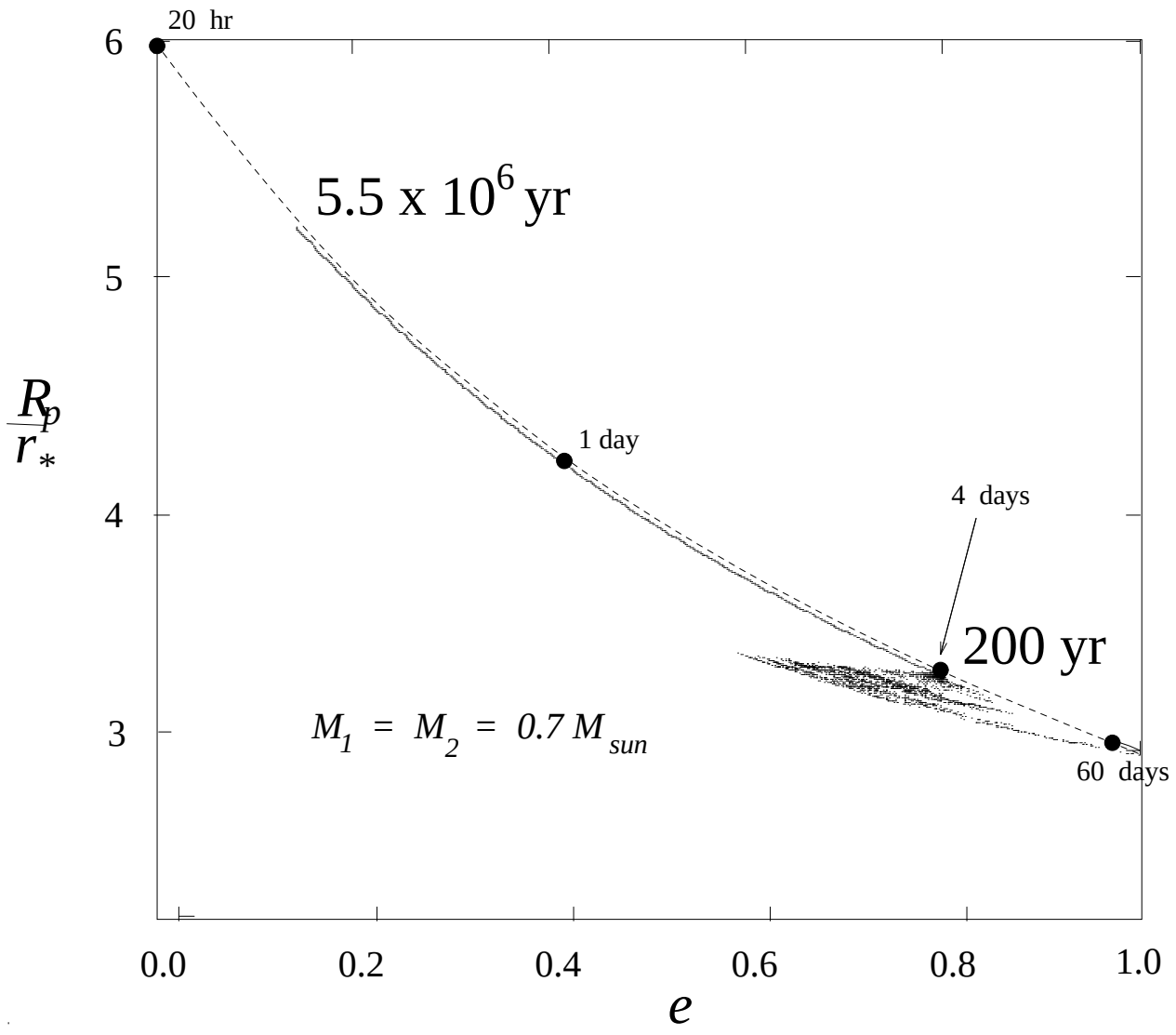

**Figure 1.12**: Chaotic and periodic circularization for the system shown in Figure 5. The times shown in days are the orbital periods at these points.

## 1.6 Discussion

It is not clear how the stars will respond to the violent tides during the chaotic phase. The stars are so close during the circularization process that any expansion in response to tidal heating will completely change the evolution. Thus it is important to know where the tidal energy is deposited. Podsiadlowski (1996) has shown that the stellar radius is not much affected if the energy is deposited near the core of the star. If the energy is deposited in the extreme outer layers, the star may lose part of these layers via a stellar wind resulting in expansion of the orbit, and this in turn will reduce the tidal interaction (although it should be noted that the chaos boundary will also be affected). On the other hand, if the energy is deposited in the intermediate layers, the effect could be devastating. Clearly the fate of tidal capture binaries depends on what happens during this phase, but one might expect a reasonable fraction (maybe 25%?) of binaries which survive the first few orbits following capture to go on and enter the quiescent phase intact.

Of those binaries which do go on to circularize "normally", how many binaries containing compact objects eventually become stable mass transfer systems? The mass ratio of a neutron star - main sequence star[12] pair favours stability against mass loss during the chaotic phase, given that results which apply to circular binaries can in some sense be carried over to highly eccentric binaries. One might in fact expect the stability criteria to be more severe, and this may explain the apparent dearth of cataclysmic variables in globular clusters (Shara 1996, but see also Grindlay 1996). It was recently suggested (Kochanek 1992) that most neutron star - main sequence star binaries formed by tidal capture in globular clusters would not survive the large angular momentum transfer involved in the circularization process. But this work assumed large tides (and hence large tidal angular momenta) all the way down to complete circularization. Given that many such binaries will spend the majority of their circularization time with rather small tides, tidal capture again becomes a viable mechanism for creating LMXBs.

Some capture binaries will suffer a relatively minor chaotic phase. For example, given the rather low velocity dispersions found in open clusters and the relatively large cross section presented by a giant star, it is possible for such stars to capture other stars at "wide" initial separations. This may save such binaries from destruction and they may well go on to become interesting objects. On the other hand, giant captures in globular clusters are likely to be violent because in order for a capture to occur, the stars must approach each other extremely closely initially.

It remains to be seen whether or not tidal capture binaries are important dynamically in globular clusters. Since many primordial binaries are destroyed in the core over the lifetime of a cluster, it may well be that, at least for very dense clusters, tidal capture binaries have an important role to play. The cluster calculations discussed above (Aarseth 1996) are presently being run on the HARP-2 machine at Cambridge with $N = 10,000$, appropriate to large open clusters. Larger simulations with $N = 25,000$ representing small globular clusters will soon be implemented on the HARP-3 machine which has recently been installed at Cambridge.

Whatever the survival fraction of tidal capture binaries is, it is clear that the two-phase behaviour revealed by the chaos model for tidal interaction changes the way we now look at the process of tidal capture.

[12]The turnoff mass of a main sequence star in a globular cluster is around $0.8M_\odot$.

## References

Aarseth, S. J.: 1974, *Astron. & Astrophys.* **35**, 237

Aarseth, S. J.: 1996, to appear in E. F. Milone and J.-C. Mermilliod (eds), *The Origins, Evolution and Destinies of Binaries in Clusters*, ASP Conference Series

Benz, W. and Hills, J. G.: 1992, *Astrophys. J.* **389**, 546

Boss, A. P.: 1985, *Comments on Ap.* **12**, 169

Carter, B. and Luminet, J. P.: 1985, *Mon. Not. R. Astron. Soc.* **212**, 23

Davies, M. B.: 1995, *Mon. Not. R. Astron. Soc.* , in press

Davies, M. B. and Benz, W.: 1995, *Mon. Not. R. Astron. Soc.* , in press

Durisen, R. H. and Tohline, J. E.: 1985, in *Protostars and Planets II*, D.C. Black and M.S. Matthews, eds., p. 534. University of Arizona Press, Tucson

Fabian, A. C., Pringle, J. E. and Rees, M. J.: 1975, *Mon. Not. R. Astron. Soc.* , **172**, 15P

Giersz, M.: 1986, *Acta Astron.*, **36**, 181

Gingold, R. A. and Monaghan, J. J.: 1980, *Mon. Not. R. Astron. Soc.* , **191**, 897

Grindlay, J. in P. Hut and J. Makino (eds.), *Dynamical Evolution of Star Clusters: Confrontation of Theory and Observations*, Kluwer Academic Publishers, Dordrecht, in press

Gunn, J. E. and Griffin, R. F.: 1979, *Astronom. J.*, **84**, 752

Hut, P., Murphy, B. W. and Verbunt, F.: 1991, *Astron. & Astrophys.*, **241**, 137

Jackson, J. D.: 1975, *Classical Electrodynamics* (New York: Wiley)

Katz, J. I.: 1975, *Nature*, **253**, 698

Kochanek, C. S.: 1992, *Astrophys. J.* **385**, 604

Lee, H. M. and Ostriker, J. P.: 1986, *Astrophys. J.* **310**, 176

Mardling, R. A.: 1991, *Chaos in Binary Star Systems*, Ph.D thesis, Monash Univ.

Mardling, R. A.: 1995a, *Astrophys. J.* , **450**, 722

Mardling, R. A.: 1995b, *Astrophys. J.* , **450**, 732

Mardling, R. A.: 1996 in P. Hut and J. Makino (eds.), *Dynamical Evolution of Star Clusters: Confrontation of Theory and Observations*, Kluwer Academic Publishers, Dordrecht, in press

McMillan, S. L. W., McDermott, P. N. and Taam, R. E.: 1987, *Astrophys. J.* , **318**, 261

Podsiadlowski, Ph.: 1996, *Mon. Not. R. Astron. Soc.* , in press

Pomeau, Y. and Manneville, P.: 1980, *Commun. Math. Phys.* **74**, 189

Press, W. H. and Teukolsky, S. A.: 1977, *ApJ*, **213**, 183

Pryor, C., McClure, R. D., Hesser, J. E. and Fletcher, J. M.: 1989, in *Dynamics of Dense Stellar Systems*, ed. D. Merritt, (New York: Cambridge University Press), 175

Rasio, F. A. and Shapiro, S. L. 1991, *Astrophys. J.* **377**, 559

Ray, A., Kembhavi, A. K. and Antia, H. M.: 1987, *Astron. & Astrophys.***184**, 164

Shara, M. M. in P. Hut and J. Makino (eds.), *Dynamical Evolution of Star Clusters: Confrontation of Theory and Observations*, Kluwer Academic Publishers, Dordrecht, in press

Shu, F.H. Adams, F.C. and Lizano, S. 1987, *Ann. Rev. Astron. Ap.* **25**, 23

Statler, T. S., Ostriker, J. P. and Cohn, H. N.: 1987, *Astrophys. J.* , **316**, 626